\documentclass{article}

\usepackage{arxiv}

\usepackage[utf8]{inputenc} % allow utf-8 input
\usepackage[T1]{fontenc}    % use 8-bit T1 fonts
\usepackage{hyperref}       % hyperlinks
\usepackage{url}            % simple URL typesetting
\usepackage{booktabs}       % professional-quality tables
\usepackage{amsfonts}       % blackboard math symbols
\usepackage{nicefrac}       % compact symbols for 1/2, etc.
\usepackage{microtype}      % microtypography
\usepackage{lipsum}
\usepackage{graphicx}
\graphicspath{ {./images/} }

\usepackage{graphicx}
\usepackage{natbib}
\usepackage{graphicx}  % For including images
\usepackage{wrapfig}   % For wrapping text around images
\usepackage{graphicx}%
\usepackage{rotating}
\usepackage{bbm}
\usepackage{bm}
\usepackage{array}  %
\usepackage{hyperref}
\usepackage{multirow}%
\usepackage{amsmath,amssymb,amsfonts}%
\usepackage{mathtools}
\usepackage{amsthm}%
\usepackage{mathrsfs}%
\usepackage[title]{appendix}%
\usepackage{xcolor}%
\usepackage{textcomp}%
\usepackage{manyfoot}%
\usepackage{booktabs}%
\usepackage{lscape}%
\usepackage{pdfpages}%
\usepackage{algorithm}%
\usepackage{algorithmicx}%
\usepackage{algpseudocode}%
\usepackage{listings}%
\usepackage{amsthm}
\newtheorem{lemma}{Lemma}

\title{Optimizing Forecast Combination Weights Using Exponentially Weighted Hit and Win Rate Losses}

\author{
 Henry D. van Eijk \\
  Department of Statistics\\
  North Carolina State University\\
  Raleigh, NC 27695 \\
  \texttt{hdvaneij@ncsu.edu} \\
   \And
 Sujit K. Ghosh \\
  Department of Statistics\\
  North Carolina State University\\
  Raleigh, NC 27695 \\
  \texttt{sujit.ghosh@ncsu.edu} \\
}

\begin{document}
\maketitle
\begin{abstract}
Forecasting revenues by aggregating analyst forecasts is a fundamental problem in financial research and practice. A key objective in this context is to improve the accuracy of the forecast by optimizing two performance metrics: the hit rate, which measures the proportion of correctly classified revenue surprise signs, and the win rate, which quantifies the proportion of individual forecasts that outperform an equally weighted consensus benchmark. While researchers have extensively studied forecast combination techniques, two critical gaps remain: (i) the estimation of optimal combination weights tailored to these specific performance metrics and (ii) the development of Bayesian methods for handling missing or incomplete analyst forecasts. This paper proposes novel approaches to address these challenges. First, we introduce a method for estimating optimal forecast combination weights using exponentially weighted hit and win rate loss functions via nonlinear programming. Second, we develop a Bayesian imputation framework that leverages exponentially weighted likelihood methods to account for missing forecasts while preserving key distributional properties. Through extensive empirical evaluations using real-world analyst forecast data, we demonstrate that our proposed methodologies yield superior predictive performance compared to traditional equally weighted and linear combination benchmarks. These findings highlight the advantages of incorporating tailored loss functions and Bayesian inference in forecast combination models, offering valuable insights for financial analysts and practitioners seeking to improve revenue prediction accuracy. 
\end{abstract}

% keywords can be removed

\section{Introduction}\label{sec1}
In the financial industry, the U.S. Securities and Exchange Commission (SEC) mandates that publicly traded companies disclose their financial statements every quarter to ensure transparency and provide investors with key insights into their financial health. Anticipating these disclosures, financial analysts generate forecasts of fundamental financial metrics—such as earnings per share (EPS) and revenue—that serve as critical indicators of a company's profitability and growth potential. In addition to these widely tracked metrics, analysts also forecast company-specific indicators, such as gross bookings for Uber or the number of video subscribers for Netflix \citep{bib11}. These forecasts are pivotal in shaping market expectations, influencing investment decisions, and guiding corporate strategies.

The financial media closely monitor these earnings announcements and assess company performance by comparing reported results against market expectations. A company either “beats” or “misses” expectations depending on whether its financial figures exceed or fall short of the forecasted consensus. The consensus forecast for a given metric, such as EPS or revenue, is typically computed as the average of all available analyst estimates. This equally weighted aggregation is a standard benchmark for evaluating forecast accuracy in the forecast combinations literature. However, equally weighting all analysts' forecasts may not always be optimal, as individual analysts differ in expertise, methodologies, and access to proprietary information. This has led to growing interest in methods that optimally combine analyst forecasts to improve predictive performance.
 
Generally, the goal of forecast combinations for a single point forecast is to optimally combine multiple forecasts to minimize some loss function $\mathcal{L}(\cdot)$. The equally weighted combination has been justified as a robust method because “(i) combination weights are equal and do not have to be estimated; (ii) simple averaging significantly reduces variance and bias by averaging out individual bias in many cases; and (iii) simple averaging should be considered when the uncertainty of weight estimation is taken into account” \citep{bib1}. However, the simple average (i.e., equally weighted combination) of heterogeneous variables is known to be suboptimal in reducing variance due to the following lemma.

\vspace{1em}
\begin{lemma}
    Let $X_1, \ldots, X_m$ be pairwise uncorrelated random variables with $E(X_j)=\mu\;\;\forall\;j$ and $Var(X_j)=\sigma_j^2$ for $j=1,2,\ldots,m$ and weights $\bm{\omega}=(\omega_1,\ldots,\omega_m)^\prime$ satisfying $\bm{\omega}\in \mathcal{S}_m=\{\bm{\omega}\succeq\bm{0}: \bm{\omega}^\prime\bm{1}_m=1\}$. Then $E(\sum_{j=1}^m \omega_j X_{j})=\mu$ and $Var(\sum_{j=1}^m \omega_j X_{j}) \geq Var(\sum_{j=1}^m \omega^*_j X_{j})$ where $w^*_j=\frac{\frac{1}{\sigma^2_j}}{\sum_{k=1}^m \frac{1}{\sigma^2_k}}$. Clearly, if the $\sigma^2_j$'s are not all equal, then $\omega^*_j\neq \frac{1}{m}$ for some $j$ and hence $Var(\sum_{j=1}^m \omega^*_j X_{j}) < Var(\bar{X}) = Var(\frac{1}{m} \sum_{j=1}^m X_j)$.
\end{lemma}
\vspace{1em}
\noindent Therefore, the variance of some optimally weighted combinations is less than the variance of the equally weighted combination. The above lemma can be extended to correlated random variables as well. E.g., if $Var(\bm{X})=\Sigma$, where $\bm{X}=(X_1,\ldots,X_m)^\prime$ is a vector of random variables with mean $E(\bm{X})=\mu\bm{1}_m$, then the optimal weight vector is obtained by solving the following quadratic programming problem: 
\begin{equation*}
    \bm{\omega}^* = \arg\min_{\bm{\omega}\in \mathcal{S}_m} \bm{\omega}^\prime\Sigma\bm{\omega} = \frac{1}{\bm{1}_m^\prime\Sigma^{-1}\bm{1}_m}\Sigma^{-1}\bm{1}_m
\end{equation*}
if $\Sigma^{-1}\bm{1}_m,\succeq\bm{0}$. Clearly, the above solution reduces to the solution given in the above lemma when $\Sigma$ is a diagonal matrix (i.e., when $X_j$'s are pairwise uncorrelated).

Among many alternative methods of combining forecasts, choosing the best forecaster model is a common method where given \textit{m} forecasters, only the “best” forecaster is considered (i.e., assigning a weight of one to them and zeros otherwise). However, previous literature has mentioned that the model does not perform well, such as \citet{bib19}, \citet{bib20}, \citet{bib21}, and \citet{bib22} \citep{bib2}. Aside from the equally weighted combination and the best forecaster model, how can one produce an optimally weighted combination? The previous literature extensively focuses on minimizing the mean squared error (MSE) as \citet{bib2} and \citet{bib3} mention the squared error loss as the typical loss function for forecast combinations. In addition, \citet{bib17} and \citet{bib18} find strong evidence in favor of minimizing MSE in forecast combinations with macroeconomic data \citep{bib2}.

In forecast combinations applications, specifically earnings and revenue forecasting, squared error loss may not be the ideal loss function. \citet{bib16} introduced the concept of post-earnings announcement drift (PEAD), which \citet{bib28} describes as “the drift of a firm’s stock price in the direction of the firm's earnings surprise over an extended period of time”. Current literature has found minimal evidence against PEAD, also known as the earnings momentum effect, and \citet{bib26} and \citet{bib27} find evidence of PEAD in both less and highly developed financial markets \citep{bib28}. Furthermore, \citet{bib4} denote the \textit{earnings torpedo} effect as the “fact that missing analysts’ forecasts [i.e., the consensus], even by small amounts, causes disproportionately large stock price declines$\ldots$ it is the simple fact of an earnings disappointment that matters for investors$\ldots$ rather than the magnitude of the disappointment”. Since forecasting whether a company will beat or miss consensus for a given metric is necessary, a loss function for binary classification may be ideal for such scenarios. To solve this problem, consider hit rate loss that aligns with \citet{bib4} where we define a \textit{hit} as correctly classifying whether the actual value beats (exceeds) or misses (fails to exceed) the analysts' consensus estimate. This definition aligns with the revenue nowcasting company AKAnomics that defines a \textit{hit} as when “both the AKAnomics estimate and the actual value came in on the \textit{same side of consensus}” (i.e., predicting the sign of the revenue surprise, predicting a beat/miss) \citep{bib6}. Furthermore, they note that “Confidence can be expressed through many metrics (e.g., historical errors, hit rates, forward returns). Our favorite metric for this purpose is the \textit{hit rate}$\ldots$ Without access to differentiated information, one shouldn’t expect a better than 50\% hit rate. Our experience shows that Sell-Side Research does no better than 50\% either$\ldots$ higher hit rate, by definition, suggests more ability to identify revenue disconnects from consensus$\ldots$ higher hit rates historically translate into higher forward weekly returns” \citep{bib6}. AKAnomics's average hit rate is 66\% using a collection of macro, industry, and company data, and it is important to note that they do not disclose their modeling approach and companies of interest \citep{bib6}.

We define win rate loss as a 0-1 loss function where a \textit{win} is when the optimally weighted forecast is closer to the actual value than the analysts' consensus estimate. A win rate greater than 50\% illustrates having a systematic edge over consensus. Win rate may not be as beneficial as the hit rate in earnings and revenue forecasting; however, many other areas of finance, such as trading, focus on how often a strategy beats a benchmark, such as an index or the market itself. In addition, market participants may find the win rate easier to grasp than other statistical error methods such as MSE. Prior literature on win rate in revenue forecasting includes \citet{bib5} that utilized a classical linear systems model with an alternative dataset, specifically a credit card transactions dataset from Second Measure Inc., to forecast quarterly revenue. This prior study from \citet{bib5} at Massachusetts Institute of Technology (MIT) inspired the creation of Fleder's nowcasting startup Covariance \citep{bib29}. The model from \citet{bib5} achieved a win rate of 57.2\% across 34 companies that remain disclosed. Although \citet{bib5} does not directly minimize win rate loss, their model still performed well.

The two previous examples justify the hit and win rates' real-world applicability in the financial markets. Although we choose to minimize hit rate loss using weighted logistic regression with exponential discounting, minimizing win rate loss is a more complex problem. The challenge in directly minimizing win rate loss, or more generally, any 0-1 valued loss function, is because of the discontinuity and non-differentiability of indicator functions. In addition, an ideal forecast combinations framework should incorporate parameter constraints and discounting, which further makes the optimization problem more challenging. Section~\ref{sec2.2.3} introduces a novel, nonlinear programming framework for forecast combinations that attempts to solve this problem by estimating the optimal weights that is based on a smooth approximation of the indicator function by a Cauchy distribution function (see Figure \ref{fig:cdf_comparison}). In addition, we also use nonlinear programming for our constrained weighted logistic regression model, which minimizes hit rate loss due to the required parameter constraints.

In the over 50-year review of forecast combinations from \citet{bib1}, they mention many techniques to find optimally weighted combinations, such as time-varying weights, nonlinear combinations, correlations among components, and cross-learning; however, there are few Bayesian approaches for forecast combinations. One of the challenges in combining analyst forecasts for revenue forecasting is handling missing data. We propose a Bayesian imputation framework that utilizes exponentially weighted likelihood methods to account for these missing forecasts, which preserves key distributional properties. Furthermore, this approach allows for a data-driven estimation of the exponential discount factor. By incorporating a discounting term into the variance of the likelihood function, we can place a prior on the discounting factor, thus eliminating the need for hyperparameter tuning. Section~\ref{sec2.2.2} covers both of these points in detail.

The main focus of this paper is to address the two main issues stated above: (i) no previous forecast combinations literature on estimating the optimal weights under exponentially weighted hit and win rate loss, and (ii) Bayesian imputation methods using exponentially weighted likelihood methods. We present our framework regarding revenue (and earnings) forecasting; however, many other applications where analysts, or more generally forecasters, release forecasts can be considered. Other applications of this framework within finance include gross domestic product (GDP), interest rates, inflation, commodity prices, currency exchange rates, and more. 

The remainder of this paper is structured as follows. Section~\ref{sec2} presents an overview of the forecast combinations problem. Section~\ref{sec2.1} introduces three loss functions for forecast combinations. Next, Section~\ref{sec2.2} covers estimation techniques including quadratic programming in Section~\ref{sec2.2.1}, Bayesian methods in Section~\ref{sec2.2.2}, and nonlinear programming in Section~\ref{sec2.2.3}. The models are evaluated based on the predictive criteria for performance evaluation in Section~\ref{sec2.3}. Section~\ref{sec3} covers empirical results for revenue forecasting, whereas Section~\ref{sec3.1} and Section~\ref{sec3.2} introduce the dataset and out-of-sample results, respectively. Lastly, Section~\ref{sec4} touches on the concluding remarks. 

\section{Forecast Combination Models}\label{sec2}
Suppose data $(y_t, \bm{x}_t)$ for $t=1,2,\ldots,T$ is observed where $y_t$ denotes the variable of interest to predict based on historical data $(y_s, \bm{x}_s)$ for indices $s\in\mathcal{T}_t$. Here, $\bm{x}_t=(x_{t,1},\ldots\,x_{t,m})^\prime$ is a vector of $m$ analyst forecast values or feature variables available before the time $t$ and $y_t$ is the variable available at time $t$. The time points $t$ can be discrete time units (e.g., hours, days, weeks) depending on the specific application. It is important to note that to predict the value of $y_{t+h}$, only training data $\mathcal{D}_t=\{(y_s, \bm{x}_s), s\in\mathcal{T}_t\}$ is accessible and $\bm{x}_{t+h}$ for any given time point $t$ and $h>0$. 

To fix ideas, suppose $L$ denotes the {\em look-back time horizon}, and $H$ denotes the {\em forecast horizon}. Then, $\mathcal{T}_t=\{t-L+1, t-L+2,\ldots,t\}$ represents the indices for training data $\mathcal{D}_t$. The goal is to predict the values of $\{y_{t+1}, y_{t+2},\ldots,y_{t+H}\}$, based on observed training sets of $\mathcal{D}_t=\{(y_s, \bm{x}_s), s\in\mathcal{T}_t\}$, for $t=L, L+1,\ldots,T-H$. The choice of look back time $L$ and forecast horizon $H$ can be based on the specific purpose of the applications. It is common to use $L=\lceil \frac{2}{3}T\rceil$ and $H=1$ to use at least two-thirds of historical data to predict the immediate next time point value of $y$. 

Any such forecast critically depends on the choice of the divergence or distance used to measure the discrepancy between the predicted values $\{\hat{y}_{t+1},\ldots,\hat{y}_{t+H}\}$ and the actual values $\{y_{t+1}, y_{t+2},\ldots,y_{t+H}\}$. We can train forecast combination models by utilizing the desired minimum value of the divergence criteria. Let $\mathcal{L}(\hat{y}_t, y_t)$ be a loss function that measures the divergence between $\hat{y}_t$ and $y_t$ at time point $t$, where $\hat{y}_t$ is obtained based on a training model using the observed data $\mathcal{D}_t$. Then the average loss for horizon time $H$ can be taken to be $\mathcal{L}_H(t) = \frac{1}{H}\sum_{h=1}^H \mathcal{L}(\hat{y}_{t+h}, y_{t+h})$. In literature, $L$ refers to the {\em rolling window size}, and we can measure the performance of the forecast values by the plot of $\mathcal{L}_H(t)$ as a function of $t$ for $t=L, L+1,\ldots, T-H$. 

Suppose there are $m$ analysts that provide forecast values $\bm{x}_t$ before time $t$. The goal is to combine these forecast values into an optimally weighted forecast using a linear function $\hat{y}_t(\bm{\omega},\omega_0)=\omega_0 + \sum_{j=1}^m \omega_j x_{t,j} = \omega_0 + \bm{\omega}^\prime\bm{x}_t$, where $\omega_0$ is the intercept representing the overall bias after combining the forecasts and $\bm{\omega}=(\omega_1,\ldots,\omega_m)^\prime$ denotes a set of nonnegative weights, satisfying $\bm{\omega}\in \mathcal{S}_m=\{\bm{\omega}\succeq\bm{0}:\bm{\omega}^\prime\bm{1}_m=1\}$; a m-dimensional simplex where $\bm{1}_m=(1,1,\ldots,1)^\prime$ is an $m\times 1$ vector of 1's. In addition, let the equally weighted combination be denoted by $\hat{y}_t(\bm{\bar{\omega}})= \bm{\bar{\omega}}^\prime\bm{x}_t$ where $\bm{\bar{\omega}}=\frac{1}{m}(1,\ldots,1)^\prime=\frac{1}{m}\bm{1}_m$ and note that it is a special case of the optimally weighted combination when $\omega_0=0$ and $\bm{\omega}=\bm{\bar{\omega}}$. We write $\hat{y}_t(\bm{\bar{\omega}})$ instead of $\hat{y}_t(\bm{\bar{\omega}},\omega_0)$ because the bias term $\omega_0=0$ for the equally weighted combination. {\em Notice that $\bm{\hat{\omega}}$ and $\hat{\omega_0}$ will change with $t$ as the training subsets $\mathcal{D}_t$ vary}, leading to the so-called time-varying parameter (TVP) estimates. However, for the sake of simplicity of notation, we may not always use $\bm{\omega}_t$ in our subsequent description when the context is clearly understood. In this case, an estimate is obtainable from 
\begin{equation*}
    ({\hat{\omega}_{0,t}(h),\bm{\hat{\omega}}_t(h)}) = \arg\min_{\substack{\bm{\omega} \in \mathcal{S}_m \\ \omega_0 \in \mathbb{R}}} \mathcal{L}(\hat{y}_{t+h}(\bm{\omega},\omega_0), y_{t+h}).
\end{equation*}
For simplicity, the results in Section~\ref{sec2} consider a horizon of one ($H=1$) and only a single fold within the rolling window cross-validation; thus, the goal is to forecast ${y}_{L+1}$ given $\bm{x}_{L+1}$ and the parameters $\hat{\omega}_0$ and $\bm{\hat{\omega}}$ estimated from $\mathcal{D}_t$ where $\mathcal{T}_L=\{1, 2,\ldots,L\}$. However, in Section~\ref{sec3}, the results are based on many folds repeating the process for $\mathcal{T}_t$, for $t=L, L+1,\ldots,T-1$.

\subsection{Loss Functions}\label{sec2.1}
Exponential discounting places greater weight on recent observations while gradually down-weighting older data. This weighting mechanism is advantageous in forecast combination methods, where past forecasts contribute to the final prediction, but their influence diminishes over time. Taking account of this aspect, a generalized objective function for forecast combinations that incorporates exponential discounting is 
\begin{equation*}
    f(\bm{\omega},\omega_0;\lambda) = \sum_{t=1}^{L} p_t(\lambda) \cdot \mathcal{L}(\hat{y}_t(\bm{\omega},\omega_0), y_{t})
\end{equation*}
with discounting factor \(\lambda\geq 0\) that controls how much weight to give to more recent observations. In this setting, one possible normalized (i.e., $p_t(\lambda)$'s sum to 1), exponential discounting function is
\begin{equation}
\label{eq:wd}
    p_t(\lambda) = \frac{e^{-\lambda(L-t)}}{\sum_{t=1}^{L} e^{-\lambda(L-t)}}=\frac{e^{-\lambda(L-t)}(1-e^{-\lambda})}{1-e^{-\lambda L}}.
\end{equation}
Moreover, note that when $\lambda=0$, all time periods (e.g., quarters) are equally weighted. Exponential discounting in forecast combinations aligns with an exponential moving average (EMA) in finance by giving more weight to recent prices and prioritizing more recent information. The main question, of course, is how much \textit{more} weight should one give to recent observations? In other words, how do we choose the value of the discounting factor $\lambda$ in practice? Section~\ref{sec2.2.1} addresses the problem of a traditional approach of hyperparameter tuning across a grid of discounting factors. In Section~\ref{sec2.2.2}, we develop a fully hierarchical Bayesian model that addresses the uncertainty of choosing $\lambda$ using a relatively uninformative prior distribution.

The most commonly used loss function for forecast combinations is the squared error loss defined as
\begin{equation*}
    \mathcal{L}_{\text{SE}}(\hat{y}_{t}(\bm{\omega},\omega_0), y_{t})=(y_t - \hat{y}_t(\bm{\omega},\omega_0))^2=\left(y_t - (\omega_0 + \sum_{j=1}^m \omega_j x_{t,j})\right)^2.
\end{equation*}
Among others, \citet{bib4} and \citet{bib6} highlighted the importance of considering the sign of forecast errors rather than the magnitude in earnings and revenue forecasting. 

We define the notions of the {\em hit rate} and {\em win rate} that are often considered more useful than average squared errors, which ignore the directionality of the forecast errors. First, we maximize the hit rate (i.e., the total number of hits) by forecasting whether the actual $y_t$ will exceed the equally weighted consensus $\hat{y}_t(\bm{\bar{\omega}})$. Let $\tilde{y}_t$ be the binary target variable defined as
\begin{equation}
\label{eq:binary_indicator}
    \tilde{y}_t = \mathbb{I}( y_t > \hat{y}_t(\bm{\bar{\omega}})) =
    \begin{cases} 
    1, & \text{if } y_t > \hat{y}_t(\bm{\bar{\omega}}) \\
    0, & \text{otherwise}
\end{cases}
\end{equation}
where the indicator $1$ denotes when the actual value $y_t$ is greater than the equally weighted consensus $\hat{y}_t(\bm{\bar{\omega}})$, and $0$ otherwise. Notice that if $\hat{\tilde{y}}_t =\mathbb{I}(\hat{y}_{t}(\bm{\omega},\omega_0)> \hat{y}_t(\bm{\bar{\omega}}))$, then hit rate is $\Pr(\hat{\tilde{y}}_t=\tilde{y}_t)$ where $\mathbb{I}(\cdot)$ is the indicator function as defined in (\ref{eq:binary_indicator}). Consider Table~\ref{tab:hit_rate_scenarios} for an illustration of a \textit{hit} which clearly shows that a hit occurs only when $\title{y}_t=\hat{\tilde{y}}_t$.

\begin{table}[ht]
\centering
\begin{tabular}{ccccc}
\toprule
Scenario & Ordering & Outcome & $\tilde{y}_t$ & $\hat{\tilde{y}}_t$ \\
\midrule
1 & $\hat{y}_{t}(\bm{\omega},\omega_0) < y_t < \hat{y}_t(\bm{\bar{\omega}})$ & Hit & 0 & 0\\
2 & $y_t < \hat{y}_{t}(\bm{\omega},\omega_0) < \hat{y}_t(\bm{\bar{\omega}})$ & Hit & 0 & 0\\
3 & $y_t < \hat{y}_t(\bm{\bar{\omega}}) < \hat{y}_{t}(\bm{\omega},\omega_0)$ & No hit & 0 & 1\\
4 & $\hat{y}_{t}(\bm{\omega},\omega_0) < \hat{y}_t(\bm{\bar{\omega}}) < y_t$ & No hit & 1 & 0\\
5 & $\hat{y}_t(\bm{\bar{\omega}}) < y_t < \hat{y}_{t}(\bm{\omega},\omega_0)$ & Hit & 1 & 1\\
6 & $\hat{y}_t(\bm{\bar{\omega}}) < \hat{y}_{t}(\bm{\omega},\omega_0) < y_t$ & Hit & 1 & 1\\
\bottomrule
\end{tabular}
\caption{Six possible scenarios for hits based on the order of the actual $y_t$, the optimally weighted forecast $\hat{y}_{t}(\bm{\omega},\omega_0)$, and the equally weighted forecast $\hat{y}_t(\bm{\bar{\omega}})$. We then illustrate how this connects to a redefined binary target variable.}
\label{tab:hit_rate_scenarios}
\end{table}

Let the the predicted probability be defined as $p_t(\bm{\omega},\omega_0)=\Pr(\tilde{y}_t=1)=[1+\exp(-\hat{y}_t(\bm{\omega},\omega_0)]^{-1}$ then we define {\bf hit rate loss} is given by
\begin{equation}
\label{eq:hr_i}
    \mathcal{L}_{\text{HR}}(\hat{p}_{t}(\bm{\omega},\omega_0), \tilde{y}_t)=-\Bigl[\tilde{y}_t \log(\hat{p}_{t}(\bm{\omega},\omega_0)) + (1 - \tilde{y}_t) \log(1 - \hat{p}_{t}(\bm{\omega},\omega_0))\Bigr]
\end{equation}
which is the negative log-likelihood for the Bernoulli distribution. Section~\ref{sec2.2.3} introduces a weighted logistic regression model with parameter constraints to minimize the loss of hit rate.

Although maximizing the hit rate can be thought of as a classification problem, a binary target variable cannot be constructed for win rate loss since it depends on our prediction $\hat{y}_t(\bm{\omega},\omega_0)$. Thus, we opt for a different approach. Let us define the relative bias as 
\begin{equation*}
    R_t(\bm{\omega},\omega_0)=\frac{y_t-\hat{y}_t(\bm{\omega},\omega_0)}{y_t-\hat{y}_t(\bm{\bar{\omega}})}=\frac{e_t(\bm{\omega},\omega_0)}{e_t(\bm{\bar{\omega}})}
\end{equation*}
thus the win rate is defined as $\Pr(\left| R_t(\bm{\omega},\omega_0) \right| < 1)$. This definition aligns with the win rate, where a win is when the optimally weighted combination is closer to the actual than the equally weighted combination. To maximize win rate (i.e., the total number of wins), let {\bf win rate loss} be defined as
\begin{equation}
\label{eq:wr_i}
    \mathcal{L}_{\text{WR}}(\hat{y}_{t}(\bm{\omega},\omega_0), y_{t})=\mathbb{I}\left( \left| R_t(\bm{\omega},\omega_0) \right| > 1 \right) 
    = \mathbb{I}\left (\left| R_t(\bm{\omega},\omega_0) \right| - 1 > 0 \right).
\end{equation}
In terms of relative bias, $\Pr(R_t(\bm{\omega},\omega_0) < 1)$ is another way to express the hit rate; however, we proceed with the definition in \eqref{eq:hr_i}. As defined above, win rate loss is a discontinuous function of $\bm{\omega}$ and $\omega_0$. Hence, it cannot be used directly from an optimization point of view to minimize the loss function. In Section~\ref{sec2.2.3}, we provide a framework to approximate this discontinuous loss function by utilizing smooth functions that enable the estimation of $\bm{\omega}$ and $\omega_0$ for optimal forecast combinations.

\subsection{Estimation Methods}\label{sec2.2}
Let $A_t$ denote the actual observed value of the forecast at time $t$ and $F_{t,j}$ be the forecast value made by the $j$-th analyst before time $t$. The actual and forecast values are positive for this application, so logarithmic values are used. Let $y_t=\log(A_t)$ and $x_{t,j}=\log(F_{t,j})$, for $t=1,2,\ldots,L$ and $j=1,2,\ldots,m$. The goal is to linearly combine the log forecast values (the $x_{t,j}$'s) to predict the log actual value (the $y_t$'s) by using $\hat{y}_t(\bm{\omega},\omega_0) = \omega_0 + \bm{\omega}^\prime\bm{x}_t$. Notice that, on the original scale, this amounts to using a {\em geometric mean} based combined forecast $\hat{A}_t(\bm{\omega},\omega_0)=\exp\{\hat{y}_t(\bm{\omega},\omega_0)\}=\exp\{\omega_0\} \prod_{j=1}^m F_{t,j}^{\omega_j}$, where $\omega_0\in\mathbb{R},\; \omega_j\geq 0,\;\forall j\geq 1$ and $\sum_{j=1}^m\omega_j=1$, i.e., $\bm{\omega}\in \mathcal{S}_m$. 

Logarithmic values are preferred because they help stabilize the variance of the errors, and the common homoscedasticity assumption can be roughly justified. For example, if the quarterly revenue of a company increases over time, logarithmic errors capture the relative error instead of the absolute error because 
\( \log\left(\frac{a}{b}\right)=\log\left(\frac{a \cdot c}{b \cdot c}\right) \) for any $c>0$. This property is not valid for absolute errors. Another useful property of logarithmic errors is the case of symmetry where \( \log\left(\frac{a}{b}\right)=-\log\left(\frac{b}{a}\right) \). 

\subsubsection{Quadratic Programming for Weighted Squared Error Loss}\label{sec2.2.1}
To obtain the optimal estimates $\bm{\hat{\omega}}$ and $\hat{\omega}_0$, we fix the discount factor $\lambda\geq 0$ to some arbitrary value and obtain the optimal parameters that minimize the chosen loss function. We repeat this optimization procedure across a grid of $\lambda$ values. The choice of $\lambda$ can be tricky, and we provide some practical justifications in our numerical illustrations (see Section~\ref{sec3}). The objective is to estimate $\bm{\omega}\in {\cal S}_m$ and $\omega_0\in\mathbb{R}$ that minimizes the discounted mean squared error given by
\begin{equation}
\begin{aligned}
\label{eq:mse}
    f(\bm{\omega},\omega_0;\lambda) 
    &=  \sum_{t=1}^{L} p_t(\lambda) \cdot \mathcal{L}_{\text{SE}}(\hat{y}_{t}(\bm{\omega},\omega_0), y_{t}) \\
    &= \sum_{t=1}^{L} \frac{e^{-\lambda(L-t)}}{\sum_{t=1}^{L} e^{-\lambda(L-t)}} \cdot (y_t - \hat{y}_t(\bm{\omega},\omega_0))^2 \\
    &= \sum_{t=1}^{L} \frac{e^{-\lambda(L-t)}}{\sum_{t=1}^{L} e^{-\lambda(L-t)}} \cdot (y_t - (\omega_0 + \sum_{j=1}^m \omega_j x_{t,j}))^2. \\
\end{aligned}
\end{equation}
Above, \eqref{eq:mse} is a constrained, weighted least squares problem where one constructs a set of linear inequality and equality constraints. To obtain the optimal weights ${\bm{\hat{\omega}}}=(\hat{\omega}_1,\ldots,\hat{\omega}_m)^\prime$ and intercept $\hat{\omega}_0$, we utilize quadratic programming, specifically, the function $\texttt{solve.QP}$ from the R package \texttt{quadprog} \citep{bib12} to address the general quadratic programming problem in the form
\begin{equation}
\begin{aligned}
\label{eq:qp}
    &\min_{{\bm{x}\in\mathbb{R}^n}} \frac{1}{2} \bm{x}^\prime \bm{D} \bm{x} - \bm{c}^\prime \bm{x} \\
    &\quad\text{s.t.} \quad \bm{A} \bm{x} \geq \bm{b}.
\end{aligned}
\end{equation}
The objective function from \eqref{eq:mse} can be expressed in quadratic form to align with \eqref{eq:qp}. Let $\bm{y} = (y_1,y_2,\ldots,y_L)^\prime$ and $\bm{X}$ is the $L\times (m+1)$ matrix whose $t$-th row consists of $(1, x_{t,1}, x_{t,2},\ldots,x_{t,m})$. Let ${\tilde{\bm{\omega}}} = (\omega_0,\bm{\omega})^\prime$ be the $(m+1)$ dimensional vector consisting of the intercept term and the $m$-dimensional weight vector parameter to be estimated, and finally let the matrix for the exponential discounting $\bm{W}=\bm{W}(\lambda)$ be an $L \times L$ diagonal matrix with $p_t(\lambda)$ for $t=1,2,\ldots,L$ from \eqref{eq:wd} as its diagonal entries. Thus, the weighted least squares objective function is
\begin{equation*}
\begin{aligned}
    \min_{\substack{\bm{\omega} \in \mathcal{S}_m \\ \omega_0 \in \mathbb{R}}} \frac{1}{2} (\bm{y} - \bm{X} \tilde{\bm{\omega}})^\prime \bm{W} (\bm{y} - \bm{X} \tilde{\bm{\omega}}) 
    &= \min_{\substack{\bm{\omega} \in \mathcal{S}_m \\ \omega_0 \in \mathbb{R}}} \frac{1}{2} \bm{y}^\prime \bm{W} \bm{y} - (\bm{y}^\prime \bm{W} \bm{X}) \tilde{\bm{\omega}} + \frac{1}{2} \tilde{\bm{\omega}}^\prime \bm{X}^\prime \bm{W} \bm{X} \tilde{\bm{\omega}} \\
    &= \min_{\substack{\bm{\omega} \in \mathcal{S}_m \\ \omega_0 \in \mathbb{R}}} \frac{1}{2} \tilde{\bm{\omega}}^\prime \bm{X}^\prime \bm{W} \bm{X} \tilde{\bm{\omega}} - (\bm{X}^\prime \bm{W} \bm{y})^\prime \tilde{\bm{\omega}} + \frac{1}{2} \bm{y}^\prime \bm{W} \bm{y} \\
    &= \min_{\substack{\bm{\omega} \in \mathcal{S}_m \\ \omega_0 \in \mathbb{R}}} \frac{1}{2} \tilde{\bm{\omega}}^\prime \bm{D} \tilde{\bm{\omega}} - \bm{c}^\prime \tilde{\bm{\omega}} + \frac{1}{2} \bm{y}^\prime \bm{W} \bm{y},
\end{aligned}
\end{equation*}
where $\bm{D} = \bm{X}^\prime \bm{W} \bm{X}$, $\bm{c} = \bm{X}^\prime\bm{W} \bm{y}$, and we can drop $\bm{y}^\prime \bm{W} \bm{y}$ because it does not depend on the parameters (for a fixed $\lambda\geq 0$). Furthermore, the constraints mentioned by $\bm{\omega} \in \mathcal{S}_m$  and $\omega_0 \in \mathbb{R}$ can be expressed as $\bm{A}\tilde{\bm{\omega}} \geq \bm{b}$. The R code available upon request on our GitHub page provides the formulation of the inequality and equality constraints matrix $\bm{A}$ and vector $\bm{b}$ from \eqref{eq:qp}. To ensure numerical convexity and stability of the quadratic term $\tilde{\bm{\omega}}^\prime \bm{D} \tilde{\bm{\omega}}$, the R function \texttt{nearPD} from the package \texttt{Matrix} \citep{bib13} is used to find the nearest positive definite matrix $\bm{D}$ ensuring that a numerically unique global minimum exists.

After estimating $\bm{\hat{\omega}}(\lambda)=(\hat{\omega}_1(\lambda),\ldots,\hat{\omega}_m(\lambda))^\prime$ and intercept $\hat{\omega}_0(\lambda)$ based on $\mathcal{T}_L=\{1, 2,\ldots,L\}$ for a given $\lambda\geq 0$, the optimally weighted combination forecast for time $L+1$ is denoted as 
\begin{equation}
\label{eq:yhatquadprog}
    \hat{y}_{L+1}(\lambda) =  \hat{\omega}_0(\lambda) + \sum_{j=1}^m \hat{\omega}_j(\lambda) x_{L+1,j}.
\end{equation}
Instead of a single combined forecast, we obtain a grid of forecast values $\hat{y}_{L+1}(\lambda)$ for a chosen grid of $\lambda$ values. Admittedly, choosing a finite set of grid values of $\lambda$ can be difficult in practice and computationally demanding if we use a finer set of grid values. This motivates us to develop a full Bayesian hierarchical model that avoids choosing a finite set of grid values and accounts for extra uncertainty introduced by the discounting factor $\lambda$.

\subsubsection{A Full Hierarchical Bayesian Framework}\label{sec2.2.2}
Clearly, the choice of $\lambda$ requires prior knowledge about the discounting rate of information over the historical data. Instead of specifying a fixed value of $\lambda$ on a given discrete grid, it is perhaps better to use a range of continuous values of $\lambda\in (c_0, d_0)$ and let the data choose the optimum value. We can overcome this limitation by utilizing a full Bayesian inferential framework. Let a Bayesian forecast combination framework that allows for exponential discounting for $\mathcal{T}_L=\{1, 2,\ldots,L\}$ be denoted as
\begin{equation}
\label{eq:bhm}
\begin{aligned}
    &y_t\sim \text{N}\left(w_0 +\sum_{j=1}^{m} x_{t,j} \omega_j, \frac{\sigma^2}{p_t(\lambda)} \right)\;
    \mbox{where}\; p_t(\lambda) = \frac{e^{-\lambda(L-t)}}{\sum_{t=1}^{L} e^{-\lambda(L-t)}},\\
    &\bm{\omega} \sim \text{Dir}(\alpha_1, \alpha_2, \ldots, \alpha_m), \\
    &w_0 \sim \text{N}(0, \tau_{w_0}^2), \\
    &\lambda \sim \text{U}(c_0, d_0), \\
    &\sigma^2 \sim \text{InvGamma}(a_0, b_0),
\end{aligned}
\end{equation}
with the goal of predicting the out-of-sample value $Y_{L+1} \sim \text{N}(w_0+\bm{\omega}^\prime \bm{{x}}_{L+1}, \sigma^2)$. In above, $\text{N}(\mu, \sigma^2)$ denotes a normal distribution with mean $\mu\in\mathbb{R}$ and variance $\sigma^2>0$, $\text{Dir}(\alpha_1, \alpha_2, \ldots, \alpha_m)$ denotes a Dirichlet distribution with parameters $\alpha_1>0, \alpha_2>0, \ldots, \alpha_m>0$, $\text{U}(c_0, d_0)$ denotes a uniform distribution with lower bound $c_0>0$ and upper bound $d_0>c_0$, and $\text{InvGamma}(a_0, b_0)$ denotes Inverse Gamma distribution with shape parameter $a_0>0$ and rate parameter $b_0>0$. Instead of utilizing hyperparameter tuning across the discounting factor $\lambda$, we specify a prior distribution for $\lambda$ to handle the uncertainty of which values to use.

Presumably, some values of $x_{t,j}$ may be missing for $t$'s where an analyst $j$ may not have provided a forecast of $a_t$. We will address the missing data model for the quadratic and nonlinear programming approaches in \ref{sec3.2}. However, this missing data problem can be more suitably handled within a Bayesian model, allowing for proper imputation of missing forecasts using the relevant posterior predictive distribution under the assumption of missing at random. Let 
\begin{equation}
\label{eq:miss_x}
\begin{aligned}
    X_{1,j} &\sim \text{N}(\gamma_j,\sigma_j^2), \;\mbox{for}\;j=1,\ldots,m\\
    X_{l+1,j}|X_{l,j}=x_{l,j}& \sim \text{N}(\gamma_j+\phi_j(x_{l,j}-\gamma_j),\sigma_j^2), \;\mbox{for}\;l=1,\ldots,L-1\\
    \phi_j &\sim \text{U}(e_0, f_0), \\
    \gamma_j &\sim \text{N}(\bar{x}_j, \tau_{\gamma}^2), \\
    \sigma_j^2 &\sim \text{InvGamma}(g_0, h_0), 
\end{aligned}
\end{equation}
denote the framework for the AR(1) missing data model where $\bar{x}_j$ denotes the mean across $\mathcal{T}_L=\{1, 2,\ldots,L\}$ for analyst $j$. In \eqref{eq:miss_x}, if one of the $m$ analysts has a missing $X_{l,j}$ for $t=l$ and $j\in \{1,2,\ldots,m\}$, then its posterior predictive distribution is used to impute the missing value given the observed ones. As such, the Monte Carlo sampling procedure (described below) can impute missing data and naturally handle the imputation uncertainty within the estimation procedure.

We implement \eqref{eq:bhm} and \eqref{eq:miss_x} using \texttt{rjags}, an interface to \texttt{JAGS} (Just Another Gibbs Sampler) \citep{bib31} that uses Markov Chain Monte Carlo (MCMC) to obtain samples from the posterior distribution of the parameters specified in the Bayesian model. Let $\bm{\theta}$ denote the collection of all parameters specified through the models described by \eqref{eq:bhm} and \eqref{eq:miss_x}. After specifying this Bayesian model in \texttt{rjags}, the MCMC provides a desired total of $S$ (possibly dependent) samples $\bm{\theta}^{(1)}, \ldots, \bm{\theta}^{(S)}$ from the posterior distribution $\bm{\theta}$ given only the fully observed data. The missing values are imputed on the fly using the posterior predictive distribution of missing values given the observed values. Since the goal is to predict $Y_{L+1}$, consider the respective posterior predictive distribution (PPD) denoted as 
\begin{equation}
\label{eq:ppd}
\begin{aligned}
    f_{PPD}(Y_{L+1} | \mathcal{D}) 
    &= \int f(Y_{L+1},\bm{\theta} | \mathcal{D})d\bm{\theta}  \\
    & = \int f(Y_{L+1} | \bm{\theta},  \mathcal{D}) p(\bm{\theta} | \mathcal{D}) d\bm{\theta}
\end{aligned}
\end{equation}
where $f(\cdot\mid\cdot)$ denotes a generic conditional density based on the conditioning arguments, $\mathcal{D}\subseteq ((y_{1},\bm{{x}}_{1}), \ldots, (y_{L},\bm{{x}}_{L}), \bm{{x}}_{L+1})$ is the observed data, and $p(\bm{\theta}| \mathcal{D})$ is the posterior density density of $\bm{\theta}$ given $\mathcal{D}$. Notice that if some $X_{t,j}$ values are missing, those are imputed using the posterior predictive distribution of $X_{t,j}$ conditional on the observed data $\mathcal{D}$ using the equation \eqref{eq:miss_x}. 

We can bypass the computation of (possibly) high-dimensional integration required to obtain the PPD in equation \eqref{eq:ppd} by using MCMC-based sampling methods (implemented using {\tt rjags}). This allows us to generate a large number of (dependent) approximate samples from the posterior distribution $p(\bm{\theta}|\mathcal{D})$, and then we reuse these samples to generate samples from the PPD. Conditional on each of the posterior samples $\bm{\theta}^{(s)}$, we generate  $Y_{L+1}^{(s)} \sim \text{N}(w_0^{(s)}+{\bm{\omega}^{(s)}}^{\prime}\bm{x}_{L+1}, {\sigma^{(s)}}^2)$ across all MCMC random samples $\bm{\theta}^{(1)}, \ldots, \bm{\theta}^{(S)}$. Then by \eqref{eq:ppd}, it follows that $Y_{L+1}^{(s)}\sim f_{PPD}(Y_{L+1} | \mathcal{D})$. Therefore, if we want to obtain the point estimate $E[Y_{L+1}|\mathcal{D}]$, we can utilize the empirical average of the sampled values $Y_{L+1}^{(s)}$'s to compute the predictive estimate
\begin{equation*}
\begin{aligned}
    \hat{y}_{L+1} = \frac{1}{S} \sum_{s=1}^S Y_{L+1}^{(s)}.
\end{aligned}
\end{equation*}
More generally, the sampling-based approach approximates the entire posterior predictive distribution of $Y_{L+1}$ conditional on the observed data $\mathcal{D}$. E.g., by using the empirical quantiles of $Y_{L+1}^{(s)}$, we can obtain interval estimates for the future value of $Y_{L+1}$. Additionally, this approach produces a single combined forecast without creating a grid of $\lambda$'s as required in \eqref{eq:yhatquadprog} from quadratic programming. Moreover, the posterior distribution of $\lambda$ conditional on the observed data $\mathcal{D}$ quantifies the uncertainty of $\lambda$. These are additional benefits of using full hierarchical Bayesian models and an associated MCMC sampling procedure that are not accessible for other classical methods described above.

\subsubsection{Nonlinear Programming}\label{sec2.2.3}

The quadratic programming and Bayesian approaches do not directly optimize hit rate loss and win rate loss. However, the reliability of MSE in forecast combinations justifies why they could still be effective. This section estimates the parameters that directly maximize hit and win rates, equivalent to minimizing hit and win rate loss.

The problem that arises when maximizing the win rate is how to handle win rate loss defined in \eqref{eq:wr_i} because of the discontinuity of the indicator functions. Naturally, we cannot use gradient-based methods to minimize the expected loss function. Thus, we can approximate the 0-1 type loss function by using a continuous and differentiable function, known as a surrogate loss function. As the indicator function only takes values 0 and 1, any cumulative distribution function (CDF) can approximate such 0-1 loss functions. The CDF is defined as \(F(z_t;z_0,\gamma)\) where \(z_t\) is where the CDF is evaluated at, \(z_0\) is the location parameter, and \(\gamma\) is the scale parameter. It is ideal to choose one that has an effective support that matches the range of values of the variable inside the indicator function. An effective support is the practical choice of the smallest interval \((a,b]\) such that for a given CDF \(F(\cdot)\), and a small tolerance \(\epsilon\), it satisfies $F(b) - F(a) = P(a < X \leq b) \geq 1 - \epsilon$. Let \textit{Z} be the random variable inside the indicator functions. Then, our goal is to determine the largest lower bound $z_{min}$ and the smallest upper bound $z_{max}$ that satisfies
\begin{equation*}
    F(z_{\text{max}}) - F(z_{\text{min}}) = P(z_{\text{min}} < Z \leq z_{\text{max}})\geq 1 - \epsilon.
\end{equation*}
In our case, $Z=\left| R(\bm{\omega},\omega_0) \right| - 1$ for win rate loss. One challenge is how to pick \(z_{\text{min}}\) and \(z_{\text{max}}\) because Z depends on the unknown parameters $\bm{\omega}$ and $\omega_0$; however, any optimally weighted forecast with an intercept of zero is inclusively between the minimum and maximum of all analysts forecasts, expressed mathematically as
\begin{equation*}
    \min(\bm{{x}}_t) \leq \bm{\omega}^\prime \bm{{x}}_t \leq \max(\bm{{x}}_t),\;\;\forall \bm{\omega}\in {\cal S}_m.
\end{equation*}
Therefore, it makes sense to empirically calculate the possible values of the relative bias $R_t(\bm{\omega},\omega_0)=\frac{y_t-\hat{y}_t(\bm{\omega},\omega_0)}{y_t-\hat{y}_t(\bm{\bar{\omega}})}=\frac{e_t(\bm{\omega},\omega_0)}{e_t(\bm{\bar{\omega}})}$ across time-periods $t=1, \ldots, L$ and analysts $1, \ldots, m$. Since there are $B=L \cdot m$ total individual forecasts, \(z_1, \ldots, z_B\) can be computed by plugging in the following into $R_t(\bm{\omega},\omega_0)$: (i) $\hat{y}_t(\bm{\omega},\omega_0)$ with $x_{t,j}$ and corresponding $y_{t}$, and (ii) $\hat{y}_t(\bm{\bar{\omega}})$. Then the empirical bounds \(z_{\text{min}}\) and \(z_{\text{max}}\) are computed by taking the minimum and maximum across \(z_1, \ldots, z_B\). After setting \(\epsilon\) and computing the empirical bounds \(z_{\text{min}}\) and \(z_{\text{max}}\), the next step is to estimate the scale parameter \(\gamma\). However, for the Cauchy and Logistic CDFs considered, the scale parameter cannot be solved algebraically; we can use numerical approximation. The R function \texttt{uniroot} can numerically approximate \(\gamma\) for the given equation
\begin{equation*}
    F(z_{\text{max}};z_0,\gamma) - F(z_{\text{min}};z_0,\gamma) - (1-\epsilon) \approx 0.
\end{equation*}
The location parameter \(z_0\) is set to zero because this is where the “activation” happens (i.e., the threshold that causes the indicator function function to take the value one). 

\begin{figure}[ht] 
  \centering
  \includegraphics[width=0.8\textwidth]{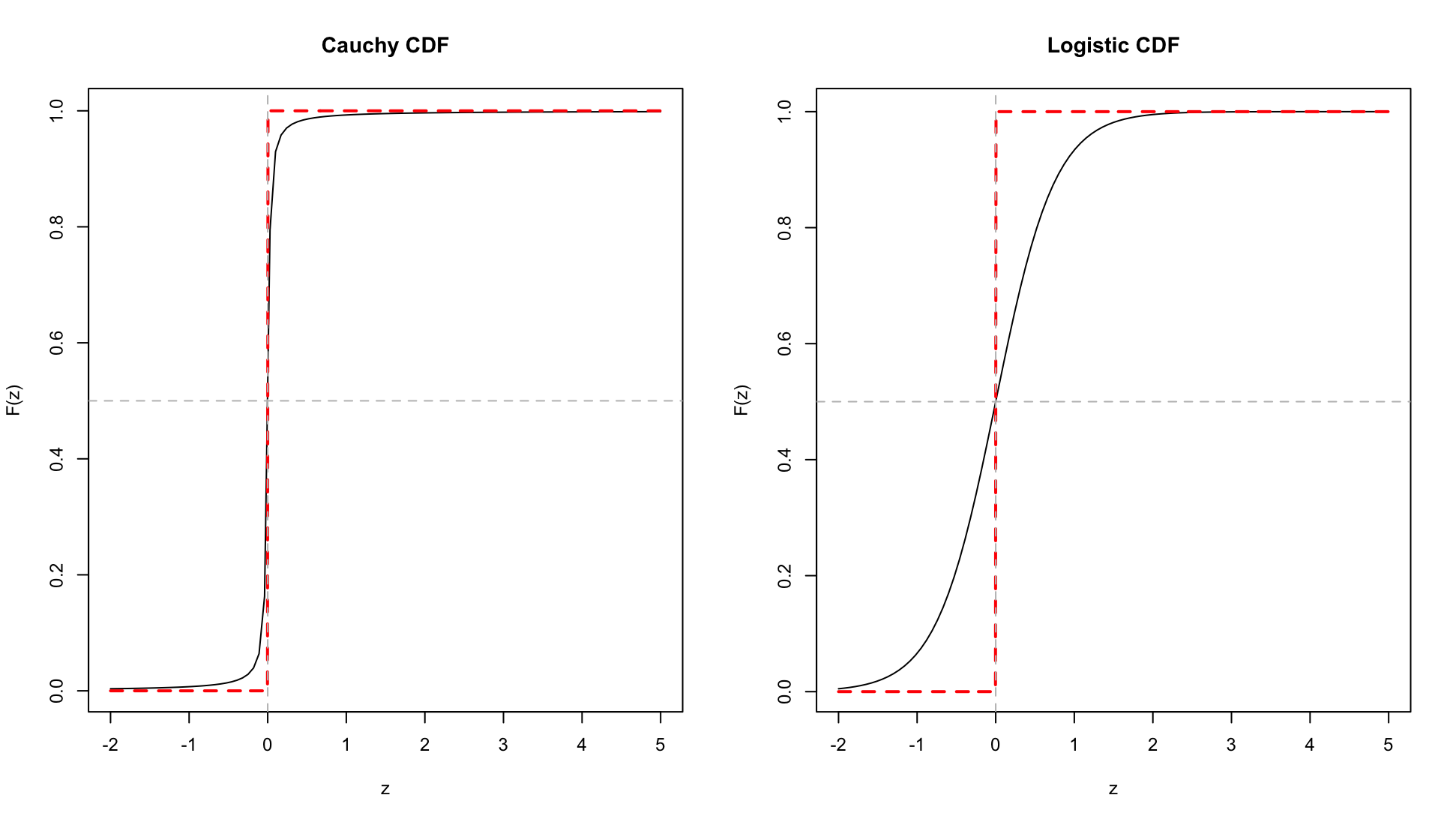} 
  \caption{Plots of the Cauchy and Logistic CDFs using the estimated scale parameter from \texttt{uniroot}. This example used a lower bound of -2, upper bound of 5, and $\epsilon=0.005$.}
  \label{fig:cdf_comparison}
\end{figure}

We define the generalized objective function to minimize discounted win rate loss as
\begin{equation*}
    f(\bm{\omega},\omega_0;\lambda) = \frac{1}{L} \sum_{t=1}^{L} p_t(\lambda) \cdot F(z_t;z_0,\hat{\gamma}) 
\end{equation*}
where \(F(z_t;z_0,\hat{\gamma})\) is a smooth approximation of the indicator function. Since \(F(z_t;z_0,\hat{\gamma})\) is a nonlinear function, we utilize nonlinear programming to obtain the optimal weights ${\bm{\hat{\omega}}}=(\hat{\omega}_1,\ldots,\hat{\omega}_m)^\prime$ and intercept $\hat{\omega}_0$. The function \texttt{nloptr} within the R package \texttt{nloptr} is utilized with the general nonlinear programming problem defined as
\begin{equation*}
\begin{array}{c}
    \min_{\substack{\bm{x} \in \mathbb{R}^n}} f(\bm{x}) \\
    \text{s.t.} \\
    g(\bm{x}) \leq 0 \\
    h(\bm{x}) = 0 \\
    \bm{x}_L \leq \bm{x} \leq \bm{x}_U
\end{array}
\end{equation*}
where $f(\cdot)$ is the objective function, $\bm{x}$ is a vector of $n$ parameters, $g(\cdot)$ and $h(\cdot)$ represent the inequality and equality constraints respectively, and $\bm{x}_L$ and $\bm{x}_U$ are the bound constraints \citep{bib14}. We use the Cauchy CDF over the Logistic CDF because it is a closer approximation to the indicator function, as shown in Figure~\ref{fig:cdf_comparison}. The gradient-free COBYLA algorithm \citep{bib25} is utilized since there are still sharp transitions in the objective function (i.e., steep gradients). 

Given $z_t(\bm{\omega},\omega_0)=\left| R_t(\bm{\omega},\omega_0) \right| - 1$, we minimize win rate loss by using the defined nonlinear programming problem (with Cauchy CDF)
\begin{equation*}
\begin{aligned}
    \min_{\substack{\bm{\omega} \in \mathcal{S}_m \\ \omega_0 \in \mathbb{R}}} \frac{1}{L} \sum_{t=1}^{L} \left(\frac{e^{-\lambda(L-t)}}{\sum_{t=1}^{L} e^{-\lambda(L-t)}}\right) \cdot \left(\frac{1}{\pi} \arctan\left(\frac{z_t - z_0}{\hat{\gamma}}\right) + \frac{1}{2}\right)
\end{aligned}
\end{equation*}
and the initial values parameter \texttt{x0} for \texttt{nloptr} is set to $\bm{{\omega}}=\frac{1}{m}\bm{1}_m$ and $\omega_0=0$. Lastly, the out-of-sample forecast $\hat{y}_{L+1}$ is calculated by plugging in $\bm{{x}}_{L+1}$, $\bm{\hat{\omega}}$, and $\hat{\omega}_0$ as in \eqref{eq:yhatquadprog}.

In order to minimize hit rate loss, all that needs to change is the loss function. We again utilize the gradient-free COBYLA algorithm with \texttt{nloptr} along with the same constraints and initial guess. We define the objective function to minimize discounted hit rate loss as 
\begin{equation}
\label{eq:nloptr_hit}
\begin{aligned}
    \min_{\substack{\bm{\omega} \in \mathcal{S}_m \\ \omega_0 \in \mathbb{R}}} \frac{1}{L} \sum_{t=1}^{L} \frac{e^{-\lambda(L-t)}}{\sum_{t=1}^{L} e^{-\lambda(L-t)}} \cdot -\Bigl[\tilde{y}_t \log(\hat{p}_{t}(\bm{\omega},\omega_0)) + (1 - \tilde{y}_t) \log(1 - \hat{p}_{t}(\bm{\omega},\omega_0))\Bigr]
\end{aligned}
\end{equation}
where $\tilde{y}_t$ is the binary indicator from \eqref{eq:binary_indicator} and $\hat{p}_t(\bm{\omega},\omega_0)=\frac{1}{1+\exp({-\hat{\eta}_t(\bm{\omega},\omega_0)})}\in(0,1)$ is the predicted probability with $\hat{\eta}_t(\bm{\omega},\omega_0)=\omega_0 + \bm{\omega}^\prime\bm{x}_t$ as the log-odds. Lastly, the out-of-sample predicted probability is defined as
\begin{equation*}
    \hat{p}_{L+1}(\lambda)=\frac{1}{1+\exp({-\hat{\eta}_t(\lambda}))}
\end{equation*}
with out-of-sample log-odds defined as
\begin{equation*}
    \hat{\eta}_{L+1}(\lambda) = \hat{\omega}_0(\lambda) + \sum_{j=1}^m \hat{\omega}_j(\lambda) x_{L+1,j}.
\end{equation*}

\subsection{Predictive Criteria for Performance Evaluation}\label{sec2.3}
In Section~\ref{sec3}, we use a real-world dataset to evaluate the out-of-sample performance for the three models: (i) quadratic programming method from Section~\ref{sec2.2.1}; (ii) Bayesian model from Section~\ref{sec2.2.2}; and (iii) nonlinear programming from Section~\ref{sec2.2.3} to optimize hit rate or win rate. For the classification problem that maximizes hit rate in \eqref{eq:nloptr_hit}, we define the out-of-sample hit rate for fold \textit{f} in the rolling-window cross-validation as
\begin{equation*}
\begin{aligned}
    \hat{\text{HR}} 
    &=\frac{1}{F}\sum_{f=1}^{F} \mathbb{I}\left(\tilde{y}_f=\mathbb{I}\left(\hat{p}_f(\bm{\omega},\omega_0)>0.5 \right)\right) 
\end{aligned}
\end{equation*}
however, the other three approaches (i.e., quadratic programming, Bayesian, and nonlinear programming for win rate) use the relative bias as they are not classification problems. The out-of-sample relative bias for fold \textit{f} in the rolling-window cross-validation is defined as
\begin{equation}
\label{eq:Rf}
    \hat{\text{R}}_f(\bm{\omega},\omega_0)=\frac{y_f-\hat{y}_f(\bm{\omega},\omega_0)}{y_f-\hat{y}_f(\bm{\bar{\omega}})}=\frac{e_f(\bm{\omega},\omega_0)}{e_f(\bm{\bar{\omega}})}.
\end{equation}
Moreover, we need to calculate \eqref{eq:Rf} as it is an input inside both the hit and win rate. Then, we denote the out-of-sample hit rate as
\begin{equation*}
\begin{aligned}
    \hat{\text{HR}}=\frac{1}{F}\sum_{f=1}^{F} \mathbb{I}\left(\hat{\text{R}}_f(\bm{\omega},\omega_0) - 1 < 0 \right)
\end{aligned}
\end{equation*}
for folds $f=1, \ldots, F$. The second criterion utilized is the out-of-sample win rate expressed as
\begin{equation*}
\begin{aligned}
    \hat{\text{WR}}=\frac{1}{F}\sum_{f=1}^{F} \mathbb{I}\left( \left| \hat{\text{R}}_f(\bm{\omega},\omega_0) \right| - 1 < 0 \right).
\end{aligned}
\end{equation*}
Section~\ref{sec3} aims to evaluate the three models' out-of-sample performance using hit rate and win rate.

\section{Empirical Results}\label{sec3}
\subsection{Data}\label{sec3.1}
The revenue data collected come from the Institutional Brokers' Estimate System (I/B/E/S) within the Wharton Research Data Services (WRDS). The sample includes the top 25 companies, by market capitalization, in the technology sector from \citet{bib10} (at the time of the research). However, we omit two companies since their history length is less than the nine years of data requirement. In theory, these large companies should be a challenging subset to forecast because \citet{bib9} note that their earnings “forecasts are more accurate and have forecast errors that are smaller than analysts’ when$\ldots$ the firm size is smaller”. For each company, the dataset includes the ground truth quarterly revenues from 2015 through 2023 and all of the respective one-quarter-ahead analyst forecasts. Since some analysts may update their original forecast or publish a new one as new information becomes available within each quarter, only an analyst's most recent forecast is considered (i.e., the forecast closest to the actual release date).

\subsection{Out-of-sample Predictive Analysis}\label{sec3.2}
In the previous section, we used only a single fold within the rolling window cross-validation for simplicity. However, our study uses nine years of quarterly data (i.e., 36 total quarters) for robustness. The rolling window cross-validation has $T-L=36-12=24$ folds (i.e., $F=24$) and a horizon of one (i.e., $H=1$), resulting in one out-of-sample prediction per fold. We choose the window size $L=12$ as \citet{bib2} mentions that “the rolling window combinations generally do best for the longest windows, i.e., $\nu$ = 9 or $\nu$ = 12”. In addition, for the quadratic and nonlinear programming methods, the hyperparameter $\lambda$ (i.e., the discount factor) is tuned across the grid $\Lambda=\{0,0.25,0.50,0.75,1\}$, depicted in Table~\ref{tab:weights}.
\begin{table}[ht]
\centering
\begin{tabular}{rrrrrr}
  \hline
\textit{t} & $\lambda=0$ & $\lambda=0.25$ & $\lambda=0.5$ & $\lambda=0.75$ & $\lambda=1$ \\ 
  \hline
1 & 0.08333 & 0.01488 & 0.00161 & 0.00014 & 0.00001 \\ 
2 & 0.08333 & 0.01911 & 0.00266 & 0.00029 & 0.00003 \\ 
3 & 0.08333 & 0.02454 & 0.00438 & 0.00062 & 0.00008 \\ 
4 & 0.08333 & 0.03150 & 0.00722 & 0.00131 & 0.00021 \\ 
5 & 0.08333 & 0.04045 & 0.01191 & 0.00277 & 0.00058 \\ 
6 & 0.08333 & 0.05194 & 0.01964 & 0.00586 & 0.00157 \\ 
7 & 0.08333 & 0.06670 & 0.03238 & 0.01241 & 0.00426 \\ 
8 & 0.08333 & 0.08564 & 0.05338 & 0.02627 & 0.01158 \\ 
9 & 0.08333 & 0.10996 & 0.08801 & 0.05562 & 0.03147 \\ 
10 & 0.08333 & 0.14119 & 0.14511 & 0.11775 & 0.08555 \\ 
11 & 0.08333 & 0.18130 & 0.23924 & 0.24927 & 0.23255 \\ 
12 & 0.08333 & 0.23279 & 0.39445 & 0.52770 & 0.63212 \\ 
   \hline
\end{tabular}
\caption{Exponential discounting (i.e., $p_t(\lambda)$ rounded to five decimal places from \eqref{eq:wd}) across the grid $\Lambda=\{0,0.25,0.50,0.75,1\}$ for $L=12$.}
\label{tab:weights}
\end{table}
Given $T$, $L$, $F$, and $H$, we denote the training subsets $\mathcal{D}_t$ as $\mathcal{D}_{12}, \ldots, \mathcal{D}_{35}$ and for each training subset, the goal is to predict the $y_{t+1}$ value given the analyst forecasts $\bm{x}_{t+1}$ that are available before $y_{t+1}$ is released. 

For each training set $\mathcal{D}_t$ within the rolling window cross-validation, our study considers an analyst if they meet the two requirements: (i) made a prediction for $y_{t+1}$ (i.e., have a present forecast in $\bm{x}_{t+1}$), and (ii) made forecasts in at least 90\% of all quarters in $\mathcal{D}_t$. The first requirement guarantees we produce a valid $\hat{y}_{t+1}$ since a corresponding analyst forecast at time $t+1$ must exist after estimating each weight $\hat{\omega}_j$. Both requirements also ensure that we exclude analysts with minimal forecasting histories—whether they have stopped forecasting entirely (e.g., due to a job change, retirement, or a shift in company coverage) or have only recently started such that there is insufficient data to estimate their weights reliably. Furthermore, this aligns with the fact that “statistical guidance articles have stated that bias is likely in analyses with more than 10\% missingness” \citep{bib7} and selecting analysts with a longer forecasting history sides with \citet{bib23} who claim that individual analyst experience increases forecast accuracy \citep{bib8}. Any remaining imputation applies only to the occasional missing forecasts from active analysts. However, our study considers all analysts when computing the equally weighted combination, and once again, only an analyst's most recent quarterly forecast is utilized. For the quadratic and nonlinear programming methods, the missing values within each $\mathcal{D}_t$ are imputed using the row mean, which is the mean across all analysts $1, \ldots, m$ for a given quarter. 

For the Bayesian model, our study specified the missing data model in \eqref{eq:miss_x}. Let the prior distributions for the parameters described in \eqref{eq:bhm} and \eqref{eq:miss_x} be
\begin{equation*}
\begin{aligned}
    &\bm{\omega} \sim \text{Dir}(1, 1, \ldots, 1), \\
    &w_0 \sim \text{N}(0, 1000), \\
    &\lambda \sim \text{U}(0, 1), \\
    &\sigma^2 \sim \text{InvGamma}(0.1, 0.1), \\
    &\phi_j \sim \text{U}(-1,1), \\
    &\gamma_j \sim \text{N}(\bar{x}_j, 100), \\
    &\sigma_j^2 \sim \text{InvGamma}(0.1, 0.1), 
\end{aligned}
\end{equation*}
for $j\in \{1,2,\ldots,m\}$. Setting $\alpha_1=\alpha_2=\ldots=\alpha_m=1$ for the parameter of weights $\bm{\omega}$ indicates that all weights $\omega_1, \omega_2, \ldots, \omega_m$ are equally likely as a $\text{Dir}(1,1,\ldots,1)$ distribution is the uniform distribution supported on the simplex ${\cal S}_m$. Furthermore, the large variances in the normal distribution and the small shape and rate parameters in the Inverse Gamma distribution are relatively uninformative priors (as the large prior variances lead to essentially uniform priors). Lastly, the Uniform prior implicitly signals equal values across the range of the discount factor $\lambda$. We used standard diagnostics to check for empirical convergence of the MCMC samples using the {\tt coda} package in R, and such diagnostics indicate no specific issues with numerical convergence. Our Bayesian models used two chains, each with a burn-in of 10,000 samples followed by 20,000 samples, leading to a total of 40,000 MCMC samples generated from the posterior distribution of the parameters.

We utilized two naive forecasting methods for benchmarking our models' performance. In chapter 3.1, \cite{bib30} mentions that the average method (i.e., the equally weighted consensus in our case) can serve as a benchmark model in forecasting. However, the win rate essentially has a built-in comparison to the equally weighted consensus, and benchmarking the hit rate against the equally weighted consensus does not make sense as the goal is to forecast whether the actual $y_t$ will be greater than or less than this quantity. This motivates us to consider two other benchmarking models from \cite{bib30}; the naive method denoted as $\hat{y}_{t+1}=y_t$ and the seasonal naive method for quarterly data denoted as $\hat{y}_{t+1}=y_{t-3}$. \cite{bib30} writes that these two naive models could be the best forecasting model available. However, they will most likely serve as benchmarks. In addition, if a new method cannot beat these benchmarks, they are not worth exploring. Thus, we will use these two simple models to benchmark our quadratic programming, nonlinear programming, and Bayesian methods and present hit and win rates. 

\begin{sidewaystable}[!p]
\centering
\small 
\begin{tabular}{l|*{6}{c}|*{6}{c}|c}
\toprule
\textbf{Ticker} & \multicolumn{6}{c|}{\textbf{Quadratic Programming}} & \multicolumn{6}{c|}{\textbf{Nonlinear Programming}} & \textbf{Bayesian} \\
 & $\lambda=0$ & $0.25$ & $0.5$ & $0.75$ & $1$ & \textbf{Mean} & $\lambda=0$ & $0.25$ & $0.5$ & $0.75$ & $1$ & \textbf{Mean} & $\alpha=1$ \\
\midrule                                         
AAPL  & 75.0 & 75.0 & 70.8 & 62.5 & 66.7 & 70.0 & 83.3 & 83.3 & 79.2 & 70.8 & 66.7 & 76.7 & 83.3 \\
ACN   & 87.5 & 87.5 & 79.2 & 79.2 & 79.2 & 82.5 & 87.5 & 87.5 & 87.5 & 75.0 & 75.0 & 82.5 & 87.5 \\
ADBE  & 91.7 & 91.7 & 91.7 & 87.5 & 87.5 & 90.0 & 91.7 & 91.7 & 91.7 & 83.3 & 83.3 & 88.3 & 91.7 \\
ADI   & 87.5 & 91.7 & 91.7 & 91.7 & 91.7 & 90.9 & 87.5 & 87.5 & 83.3 & 75.0 & 75.0 & 81.7 & 87.5 \\
ADP   & 70.8 & 75.0 & 75.0 & 79.2 & 79.2 & 75.8 & 66.7 & 75.0 & 79.2 & 83.3 & 83.3 & 77.5 & 75.0 \\
AMAT  & 79.2 & 79.2 & 79.2 & 79.2 & 79.2 & 79.2 & 87.5 & 87.5 & 83.3 & 75.0 & 75.0 & 81.7 & 87.5 \\
AMD   & 75.0 & 75.0 & 75.0 & 75.0 & 62.5 & 72.5 & 79.2 & 75.0 & 75.0 & 75.0 & 75.0 & 75.8 & 79.2 \\
ANET  & 95.8 & 95.8 & 95.8 & 95.8 & 95.8 & 95.8 & 100.0 & 100.0 & 100.0 & 100.0 & 100.0 & 100.0 & 100.0 \\
AVGO  & 79.2 & 75.0 & 75.0 & 83.3 & 83.3 & 79.2 & 75.0 & 70.8 & 75.0 & 70.8 & 70.8 & 72.5 & 66.7 \\
CRM   & 100.0 & 100.0 & 100.0 & 100.0 & 100.0 & 100.0 & 100.0 & 100.0 & 100.0 & 100.0 & 100.0 & 100.0 & 100.0 \\
CSCO  & 87.5 & 87.5 & 87.5 & 87.5 & 87.5 & 87.5 & 91.7 & 91.7 & 87.5 & 83.3 & 83.3 & 87.5 & 83.3 \\
IBM   & 41.7 & 54.2 & 58.3 & 58.3 & 62.5 & 55.0 & 54.2 & 50.0 & 41.7 & 41.7 & 41.7 & 45.9 & 58.3 \\
INTC  & 79.2 & 79.2 & 79.2 & 75.0 & 79.2 & 78.4 & 83.3 & 79.2 & 75.0 & 66.7 & 66.7 & 74.2 & 83.3 \\
INTU  & 79.2 & 87.5 & 83.3 & 83.3 & 66.7 & 80.0 & 79.2 & 79.2 & 70.8 & 66.7 & 66.7 & 72.5 & 75.0 \\
KLAC  & 100.0 & 100.0 & 100.0 & 100.0 & 100.0 & 100.0 & 100.0 & 100.0 & 100.0 & 100.0 & 100.0 & 100.0 & 100.0 \\
MSFT  & 83.3 & 87.5 & 83.3 & 83.3 & 83.3 & 84.14 & 87.5 & 87.5 & 83.3 & 75.0 & 75.0 & 81.7 & 87.5 \\
MU    & 70.8 & 79.2 & 75.0 & 75.0 & 79.2 & 75.8 & 83.3 & 83.3 & 87.5 & 83.3 & 83.3 & 84.14 & 83.3 \\
NOW   & 87.5 & 87.5 & 83.3 & 83.3 & 83.3 & 85.0 & 87.5 & 87.5 & 83.3 & 87.5 & 83.3 & 85.8 & 87.5 \\
NVDA  & 79.2 & 75.0 & 75.0 & 75.0 & 75.0 & 75.8 & 87.5 & 87.5 & 83.3 & 83.3 & 83.3 & 85.0 & 87.5 \\
ORCL  & 54.2 & 50.0 & 54.2 & 58.3 & 54.2 & 54.2 & 58.3 & 50.0 & 45.8 & 37.5 & 37.5 & 45.8 & 50.0 \\
PANW  & 87.5 & 87.5 & 87.5 & 87.5 & 83.3 & 86.7 & 91.7 & 91.7 & 91.7 & 83.3 & 83.3 & 88.3 & 91.7 \\
QCOM  & 70.8 & 62.5 & 62.5 & 66.7 & 66.7 & 65.8 & 75.0 & 70.8 & 62.5 & 50.0 & 50.0 & 61.7 & 75.0 \\
TXN   & 70.8 & 70.8 & 70.8 & 66.7 & 66.7 & 69.2 & 75.0 & 75.0 & 66.7 & 70.8 & 70.8 & 71.7 & 75.0 \\
\midrule
\textbf{Mean} & 79.7 & 80.6 & 79.7 & 79.7 & 78.8 & 79.7 & 83.2 & 82.2 & 79.7 & 75.5 & 75.2 & 79.2 & 82.4 \\
\bottomrule
\end{tabular}
\caption{Hit rates (rounded to one decimal place) for the quadratic programming, nonlinear programming, and Bayesian approaches.} 
\label{tab:Hit_Rates}
\end{sidewaystable}

Table~\ref{tab:Hit_Rates} displays the hit rates across the quadratic programming, nonlinear programming, and Bayesian models. First, the quadratic programming model achieved a 79.7\% hit rate averaging over the five hyperparameter choices. Next, again averaging over hyperparameters, the nonlinear programming approach recorded a hit rate of 79.2\%. Although the nonlinear programming method is the only approach that directly minimizes hit rate loss, the 82.4\% mean hit rate from the Bayesian model achieved the best performance; however, the nonlinear programming approach with no exponential discounting (i.e., $\lambda=0$) had the best results with a hit rate of 83.2\%. The Bayesian model could have performed the best due to its ability to handle the parameter uncertainty of the discounting factor $\lambda$ more effectively than the quadratic and nonlinear programming approaches. All three approaches exceeded the naive and seasonal naive methods that posted hit rates of 39.7\% and 28.1\%, respectively.

Our literature review found no prior forecast combinations research on maximizing hit rate in earnings and revenue forecasting except for the revenue nowcasting company AKAnomics, which reports that their average hit rate is 66\% using a collection of macro, industry, and company data \citep{bib6}. In addition, they mention that Wall Street sell-side research does not do better than a 50\% hit rate \citep{bib6}. Our average hit rates of 79.7\%, 79.2\%, and 82.4\% across a robust test set of 23 companies and 24 quarters outperformed AKAnomics's 66\% hit rate. It is difficult to answer why our methods outperformed AKAnomics because their modeling approach is inaccessible to the public.  

\begin{sidewaystable}[!p]
\centering
\small 
\begin{tabular}{l|*{6}{c}|*{6}{c}|c}
\toprule
\textbf{Ticker} & \multicolumn{6}{c|}{\textbf{Quadratic Programming}} & \multicolumn{6}{c|}{\textbf{Nonlinear Programming}} & \textbf{Bayesian} \\
 & $\lambda=0$ & $0.25$ & $0.5$ & $0.75$ & $1$ & \textbf{Mean} & $\lambda=0$ & $0.25$ & $0.5$ & $0.75$ & $1$ & \textbf{Mean} & $\alpha=1$ \\
\midrule                    
AAPL  & 54.2 & 50.0 & 58.3 & 54.2 & 58.3 & 55.0 & 66.7 & 66.7 & 54.2 & 62.5 & 62.5 & 62.5 & 58.3 \\
ACN   & 54.2 & 58.3 & 54.2 & 58.3 & 50.0 & 55.0 & 79.2 & 79.2 & 62.5 & 62.5 & 66.7 & 70.0 & 62.5 \\
ADBE  & 62.5 & 62.5 & 70.8 & 66.7 & 70.8 & 66.7 & 83.3 & 83.3 & 83.3 & 79.2 & 75.0 & 80.8 & 66.7 \\
ADI   & 66.7 & 75.0 & 75.0 & 70.8 & 70.8 & 71.7 & 75.0 & 79.2 & 75.0 & 75.0 & 70.8 & 75.0 & 58.3 \\
ADP   & 54.2 & 58.3 & 62.5 & 58.3 & 58.3 & 58.3 & 58.3 & 62.5 & 66.7 & 66.7 & 70.8 & 65.0 & 62.5 \\
AMAT  & 75.0 & 70.8 & 70.8 & 66.7 & 62.5 & 69.2 & 91.7 & 87.5 & 79.2 & 75.0 & 62.5 & 79.2 & 79.2 \\
AMD   & 50.0 & 58.3 & 62.5 & 66.7 & 54.2 & 58.3 & 70.8 & 58.3 & 70.8 & 70.8 & 62.5 & 66.6 & 41.7 \\
ANET  & 62.5 & 62.5 & 58.3 & 66.7 & 70.8 & 64.2 & 70.8 & 70.8 & 75.0 & 79.2 & 79.2 & 75.0 & 66.7 \\
AVGO  & 45.8 & 50.0 & 45.8 & 54.2 & 58.3 & 50.8 & 79.2 & 83.3 & 83.3 & 83.3 & 79.2 & 81.7 & 41.7 \\
CRM   & 66.7 & 66.7 & 62.5 & 62.5 & 66.7 & 65.0 & 83.3 & 83.3 & 83.3 & 91.7 & 83.3 & 85.0 & 75.0 \\
CSCO  & 75.0 & 66.7 & 66.7 & 66.7 & 66.7 & 68.4 & 83.3 & 83.3 & 75.0 & 70.8 & 70.8 & 76.6 & 66.7 \\
IBM   & 33.3 & 41.7 & 45.8 & 37.5 & 45.8 & 40.8 & 41.7 & 37.5 & 41.7 & 45.8 & 45.8 & 42.5 & 50.0 \\
INTC  & 62.5 & 62.5 & 58.3 & 54.2 & 58.3 & 59.2 & 70.8 & 62.5 & 62.5 & 58.3 & 54.2 & 61.7 & 66.7 \\
INTU  & 70.8 & 75.0 & 70.8 & 75.0 & 58.3 & 70.0 & 79.2 & 70.8 & 66.7 & 58.3 & 58.3 & 66.7 & 62.5 \\
KLAC  & 83.3 & 83.3 & 83.3 & 87.5 & 87.5 & 85.0 & 95.8 & 95.8 & 91.7 & 91.7 & 87.5 & 92.5 & 83.3 \\
MSFT  & 75.0 & 79.2 & 75.0 & 75.0 & 70.8 & 75.0 & 83.3 & 79.2 & 75.0 & 70.8 & 70.8 & 75.8 & 75.0 \\
MU    & 45.8 & 58.3 & 50.0 & 54.2 & 58.3 & 53.3 & 54.2 & 54.2 & 58.3 & 50.0 & 62.5 & 55.8 & 58.3 \\
NOW   & 66.7 & 66.7 & 62.5 & 66.7 & 66.7 & 65.9 & 66.7 & 66.7 & 62.5 & 62.5 & 66.7 & 65.0 & 66.7 \\
NVDA  & 62.5 & 62.5 & 70.8 & 70.8 & 66.7 & 66.7 & 79.2 & 79.2 & 75.0 & 75.0 & 75.0 & 76.7 & 75.0 \\
ORCL  & 45.8 & 29.2 & 33.3 & 37.5 & 29.2 & 35.0 & 33.3 & 45.8 & 37.5 & 29.2 & 25.0 & 34.2 & 37.5 \\
PANW  & 62.5 & 62.5 & 62.5 & 66.7 & 62.5 & 63.3 & 79.2 & 87.5 & 79.2 & 75.0 & 70.8 & 78.3 & 66.7 \\
QCOM  & 58.3 & 50.0 & 50.0 & 54.2 & 54.2 & 53.3 & 66.7 & 66.7 & 54.2 & 50.0 & 50.0 & 57.5 & 62.5 \\
TXN   & 58.3 & 58.3 & 58.3 & 54.2 & 54.2 & 56.7 & 62.5 & 62.5 & 54.2 & 54.2 & 50.0 & 56.7 & 58.3 \\
\midrule
\textbf{Mean} & 60.5 & 61.2 & 61.2 & 62.0 & 60.9 & 61.2 & 71.9 & 71.6 & 68.1 & 66.8 & 65.2 & 68.7 & 62.7 \\
\bottomrule
\end{tabular}
\caption{Win rates (rounded to one decimal place) for the quadratic programming, nonlinear programming, and Bayesian approaches.}
\label{tab:Win_Rates}
\end{sidewaystable}

Win rate is the second metric used to validate our out-of-sample performance. Table~\ref{tab:Win_Rates} illustrates the win rates across all three models. We average over the hyperparameter uncertainty to obtain a win rate of 61.2\% and 68.7\%  for the quadratic and nonlinear programming approach, respectively. In addition, the Bayesian approach recorded an average win rate of 62.7\%. The best-case scenario across all models and hyperparameters is the nonlinear programming approach with a mean win rate of 71.9\% when no exponential discounting is used (i.e., $\lambda=0$). The nonlinear programming approach most likely performs the best in maximizing win rate because it is the only approach that directly minimizes win rate loss. Lastly, all three methods surpass the naive and seasonal naive methods that posted win rates of 17.0\% and 6.7\%, respectively.

\citet{bib5} utilized a classical linear systems model to forecast quarterly revenue from a sample of 34 companies from 2015 to 2019. Instead of using analyst estimates, this approach used an alternative dataset, specifically a credit card transactions dataset from Second Measure Inc. Their model achieved a win rate of 57.2\% across 306 out-of-sample predictions, while our three approaches achieved win rates of 61.2\%, 68.7\%, and 62.7\% across 552 out-of-sample predictions per model. Lastly, one limitation from \citet{bib5} is that their study did not occur during a period of market stress, such as the COVID-19 pandemic.  

Since both the hit rate and win rate depend on the consensus, directly utilizing individual analyst forecasts could be why our model outperformed prior work. It is unknown if AKAnomics's utilized analyst forecasts; however, \citet{bib5} only used alternative data, specifically a credit card transactions dataset. A direct comparison of our work against AKAnomics and \citet{bib5} is difficult because they did not disclose the companies that compose their average hit and win rates, respectively. 

Although comparing our methods against these two sources is difficult, Table~\ref{tab:CSCO_weights} gives an example of how the estimated weights $\bm{\hat{\omega}}$ and  $\hat{\omega}_0$ across each of our models vary for the ticker CSCO in the 24th fold of the rolling window cross-validation. Our previous results show that both nonlinear programming methods and the Bayesian model produced weights close to equal allocation of the weights. However, the quadratic programming approach usually had a couple of weights dominate with the rest zero. These likely illustrate why quadratic programming did not outperform the other methods by choosing forecast values from a small subset of analysts.  

\begin{sidewaystable}[!p]
\centering
\begin{tabular}{l c c c c c c c c c c c c}
\toprule
\textbf{Method} & $\lambda$
  & $\hat{\omega}_0$ & $\hat{\omega}_1$ & $\hat{\omega}_2$ & $\hat{\omega}_3$
  & $\hat{\omega}_4$ & $\hat{\omega}_5$ & $\hat{\omega}_6$ & $\hat{\omega}_7$
  & $\hat{\omega}_8$ & $\hat{\omega}_9$ & $\hat{\omega}_{10}$ \\
\midrule
\textbf{Bayesian} & --
  & 0.008 & 0.095 & 0.100 & 0.099 & 0.097 & 0.101 & 0.101 & 0.098 & 0.104 & 0.100 & 0.103 \\

% QP block
\multirow{5}{*}{\textbf{QP}}
 & 0.00 
  & 0.005 & 0.000 & 0.000 & 0.000 & 1.000 & 0.000 & 0.000 & 0.000 & 0.000 & 0.000 & 0.000 \\
 & 0.25 
  & 0.009 & 0.000 & 0.000 & 0.000 & 1.000 & 0.000 & 0.000 & 0.000 & 0.000 & 0.000 & 0.000 \\
 & 0.50 
  & 0.011 & 0.000 & 0.000 & 0.000 & 1.000 & 0.000 & 0.000 & 0.000 & 0.000 & 0.000 & 0.000 \\
 & 0.75 
  & 0.013 & 0.000 & 0.000 & 0.000 & 0.651 & 0.000 & 0.264 & 0.000 & 0.085 & 0.000 & 0.000 \\
 & 1.00 
  & 0.013 & 0.000 & 0.000 & 0.032 & 0.352 & 0.030 & 0.304 & 0.000 & 0.283 & 0.000 & 0.000 \\

% NLP--Hit block
\multirow{5}{*}{\textbf{NLP (Hit rate)}}
 & 0.00 
  & -7.868 & 0.117 & 0.084 & 0.100 & 0.188 & 0.063 & 0.097 & 0.089 & 0.078 & 0.104 & 0.080 \\
 & 0.25 
  & -7.405 & 0.199 & 0.161 & 0.021 & 0.184 & 0.368 & 0.000 & 0.000 & 0.000 & 0.000 & 0.067 \\
 & 0.50 
  & -6.487 & 0.131 & 0.145 & 0.112 & 0.109 & 0.169 & 0.059 & 0.000 & 0.032 & 0.100 & 0.143 \\
 & 0.75 
  & -5.406 & 0.162 & 0.206 & 0.079 & 0.008 & 0.199 & 0.000 & 0.039 & 0.000 & 0.089 & 0.218 \\
 & 1.00 
  & -4.265 & 0.167 & 0.134 & 0.123 & 0.063 & 0.169 & 0.039 & 0.000 & 0.000 & 0.096 & 0.207 \\

% NLP--Win block
\multirow{5}{*}{\textbf{NLP (Win rate)}}
 & 0.00 
  & 0.008 & 0.105 & 0.099 & 0.098 & 0.098 & 0.100 & 0.100 & 0.105 & 0.102 & 0.098 & 0.096 \\
 & 0.25 
  & 0.010 & 0.100 & 0.100 & 0.099 & 0.104 & 0.104 & 0.099 & 0.099 & 0.104 & 0.096 & 0.095 \\
 & 0.50 
  & 0.011 & 0.118 & 0.100 & 0.098 & 0.094 & 0.099 & 0.099 & 0.100 & 0.097 & 0.099 & 0.096 \\
 & 0.75 
  & 0.011 & 0.100 & 0.104 & 0.095 & 0.102 & 0.100 & 0.099 & 0.100 & 0.102 & 0.097 & 0.100 \\
 & 1.00 
  & 0.011 & 0.101 & 0.102 & 0.104 & 0.102 & 0.097 & 0.109 & 0.082 & 0.111 & 0.098 & 0.095 \\
\bottomrule
\end{tabular}
\caption{Comparison of estimated weights $\omega_0, \omega_1, \dots, \omega_{10}$ (rounded to three decimal places) across our Bayesian, quadratic programming, and nonlinear programming (that maximized both hit rate and win rate) methods. This example is for the ticker CSCO in the 24th fold of our rolling window cross-validation. The quadratic and nonlinear programming methods were evaluated across $\Lambda=\{0,0.25,0.50,0.75,1\}$. {\em Notice that the estimated intercepts for the hit rate model via logistic regression are largely negative. Since the parameters are in terms of log-odds, these estimates are not representative of usual predictive bias.}}
\label{tab:CSCO_weights}
\end{sidewaystable}

One possible drawback of our study is that we evaluated our out-of-sample performance against 23 large (by market capitalization) technology companies. This approach is similar to 34 companies from \citet{bib5} who also focused on a single sector, but shy of AKAnomic's 120 companies tracked \citep{bib6}. In addition, we derive our results based on quadratic and nonlinear programming (which use several values of $\lambda$ on a grid) from a research setting as, in practice, we would need to release our prediction based on a specific value of $\lambda$. In contrast, AKAnomics's and our Bayesian method (which estimates $\lambda$ based on the training data) likely reflect real-life performance. We have not yet determined the performance across different-sized companies and other sectors. However, in theory, larger firms such as the ones utilized in our study should be more difficult to forecast than smaller firms since it aligns with the claim from \citet{bib9} that their earnings forecasts are more accurate when firm size is smaller. Furthermore, they should be more difficult to forecast due to the following characteristics of larger firms: (i) less volatility, (ii) more analyst coverage, and (iii) more publicly available information.  

\begin{figure}[ht] 
  \centering
  \includegraphics[width=0.8\textwidth]{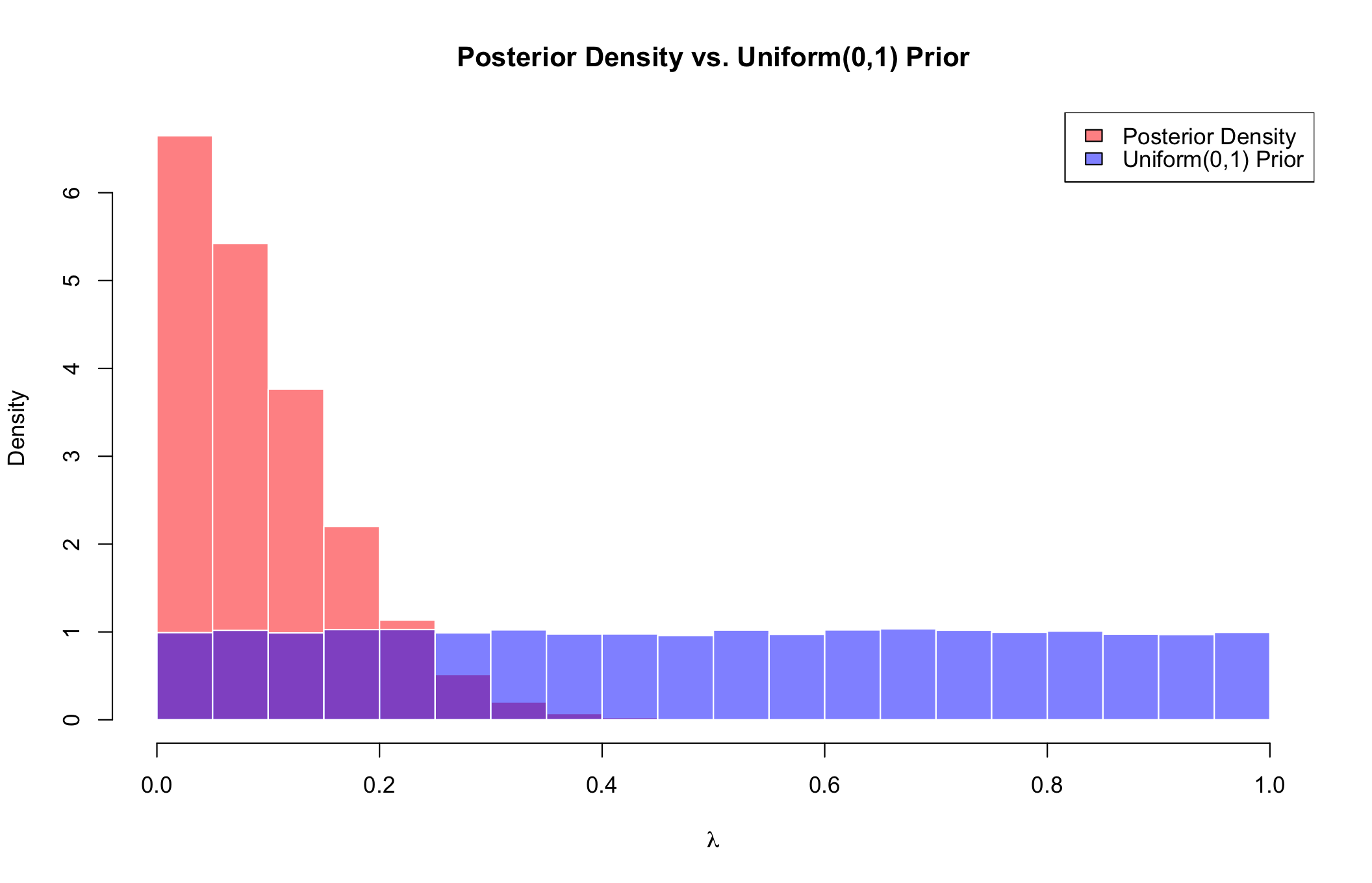} 
  \caption{Histogram of the 40,000 posterior samples of the discount factor $\lambda$ overlaid with 40,000 random samples from a U(0,1) (i.e., the prior distribution). This example is for the ticker NVDA in the 24th fold of the rolling window cross-validation.}
  \label{fig:lambdanvda}
\end{figure}

Aside from hit and win rates, one of the major takeaways from our research is the weak evidence of using exponential discounting in revenue forecasting. Table~\ref{tab:tvp} displays the average posterior mean, median, and standard deviation for $\lambda$ across all 24 folds for each ticker. Although the table presents a mean of 0.10 across all tickers, the posterior densities of $\lambda$ are highly skewed right, and the mode is approximately zero, indicating that setting $\lambda=0$ would possibly provide the best forecasts for hit and win rates. Figure~\ref{fig:lambdanvda} illustrates an example posterior histogram plot of $\lambda$ (based on MCMC samples) from a particular fold for the ticker NVDA with an overlaid uniform prior with a lower bound of zero and an upper bound of one. These results indicate weak evidence for exponential discounting for the Bayesian approach that aligns with the strong performances of small discounting factors for the quadratic and nonlinear programming approaches. One may suggest increasing the window size; however, increasing the long window size of 12 indicated by \citet{bib2} comes with a trade-off; as the window size increases, more data becomes available, but the number of forecasters ultimately decreases because the amount of missing data per forecaster increases. Nonetheless, our proposed Bayesian method provides a data-driven estimation of the discounting parameter $\lambda$, which could benefit other markets and tickers.

\begin{table}[ht]
\centering
\begin{tabular}{lccc}
\hline
\textbf{Ticker} & \textbf{Posterior mean} & \textbf{Posterior median} & \textbf{Posterior standard deviation} \\
\hline
AAPL & 0.0985 & 0.0823 & 0.0761 \\
MSFT & 0.0983 & 0.0821 & 0.0760 \\
NVDA & 0.0984 & 0.0821 & 0.0762 \\
AVGO & 0.0982 & 0.0820 & 0.0760 \\
ORCL & 0.0982 & 0.0819 & 0.0758 \\
CRM  & 0.0983 & 0.0819 & 0.0760 \\
AMD  & 0.0985 & 0.0822 & 0.0760 \\
ACN  & 0.0984 & 0.0821 & 0.0761 \\
ADBE & 0.0984 & 0.0821 & 0.0759 \\
CSCO & 0.0984 & 0.0821 & 0.0759 \\
IBM  & 0.0983 & 0.0819 & 0.0760 \\
QCOM & 0.0983 & 0.0820 & 0.0759 \\
TXN  & 0.0982 & 0.0821 & 0.0758 \\
NOW  & 0.0985 & 0.0822 & 0.0762 \\
INTU & 0.0985 & 0.0823 & 0.0762 \\
AMAT & 0.0982 & 0.0820 & 0.0759 \\
ANET & 0.0984 & 0.0821 & 0.0761 \\
ADP  & 0.0983 & 0.0821 & 0.0759 \\
MU   & 0.0984 & 0.0823 & 0.0758 \\
ADI  & 0.0986 & 0.0823 & 0.0761 \\
PANW & 0.0984 & 0.0822 & 0.0760 \\
KLAC & 0.0985 & 0.0823 & 0.0761 \\
INTC & 0.0984 & 0.0821 & 0.0761 \\
\hline
\end{tabular}
\caption{Average posterior mean, median, and standard deviation (rounded to four decimal places) for the discount factor $\lambda$ across all 24 folds for each ticker.}
\label{tab:tvp}
\end{table}

\section{Discussion}\label{sec4}
Our paper aims to address two key gaps in forecast combinations literature: (i) exponentially discounted hit and win rate losses using nonlinear programming to estimate optimal parameters and (ii) Bayesian imputation approaches using exponentially weighted likelihood methods. We evaluate our models' out-of-sample performance against the real-world revenue nowcasting company AKAnomics and the research from \citet{bib5} at Massachusetts Institute of Technology (MIT). Our quadratic programming, nonlinear programming, and Bayesian models achieved hit rates of 79.7\%, 79.2\%, and 82.4\% exceeding AKAnomic's 66\% hit rate. In addition, our models recorded win rates of 61.2\%, 68.7\%, and 62.7\% outperforming the 57.2\% win rate from \citet{bib5}. We display the robustness of our results by utilizing a 24-quarter (i.e., six-year) test set through COVID-19, a time of market uncertainty.

Further research could include the following: (i) methods to convert minimized loss function approach to Bayesian likelihood-based models (e.g., using Gibbs distributions), (ii) including other features and predictors in our models, such as macroeconomic or alternative data sources, and (iii) utilizing other surrogate loss functions in our nonlinear programming approach to approximate 0-1 type loss functions. Given that losses are not necessarily symmetric regarding forecasting, we may also consider other asymmetric loss functions that account for the asymmetry of forecast errors. Our code (using R software) will be available on GitHub following the publication of our research.

\bibliographystyle{unsrtnat}  
\bibliography{references}

\end{document}